\documentclass[preprint,12pt]{elsarticle}

\usepackage{amssymb}
\usepackage{amsmath}
\usepackage[numbers]{natbib}
\usepackage{xcolor}
\usepackage{makecell}
\usepackage{array,booktabs,makecell}
\usepackage{siunitx}
\usepackage{lineno}
\usepackage{placeins}
\usepackage{hyperref}
\usepackage{graphicx}
\usepackage{caption}
\usepackage{subcaption}
\usepackage{tabularx}
\usepackage{soul}
\usepackage{xcolor}

\journal{Particuology}

\begin{document}

\begin{frontmatter}

\title{Identifying Optimal Regression Models For DEM Simulation Datasets}

\author[add1,add2]{B.D. Jenkins\corref{cor}}
\author[add1]{A.L. Nicu\c{s}an}
\author[add2]{A. Neveu}
\author[add3]{G. Lumay}
\author[add2]{F. Francqui}
\author[add1]{J.P.K. Seville}
\author[add1]{C.R.K. Windows-Yule}
\cortext[cor]{bdj746@student.bham.ac.uk}

\address[add1]{School of Chemical Engineering, the University of Birmingham, Edgbaston, Birmingham, B15 2TT, UK}
\address[add2]{Granutools, Rue Jean Lambert Defr$\mathrm{\hat{e}}$ne 107, 4340 Awans, Belgium}
\address[add3]{Grasp laboratory, CESAM research unit, University of Liège, Place du 20 Août 7, 4000 Liège, Belgium}

\begin{abstract}

Developing fast regression models (surrogate/metamodels) from DEM data is key for practical industrial application to allow real-time evaluations. However, benchmarking different models is often overlooked in particle technology for regression tasks, as model selection is frequently not the primary research focus. This can lead to the use of suboptimal models, resulting in subpar predictive accuracy, slow evaluations, or poor generalisation, hindering effective real-time decision-making and process optimisation. In this work, we discuss applying k-fold cross-validation to assess regression models for tabular DEM datasets and propose a simple framework for readers to follow to find the optimal model for their data. An example demonstrates its application to a DEM dataset of packing fractions measured in a simple measuring beaker with varying inter-particle properties, namely, average particle diameter, coefficient of restitution, coefficient of sliding friction, coefficient of rolling resistance, and cohesive energy density. Out of 16 different models tested, a histogram-based gradient boosting model was found to be optimal, providing a good fit with acceptable training and inference times.

\end{abstract}

\begin{graphicalabstract}
\begin{figure}
    \centering
    \includegraphics[width=0.98\linewidth]{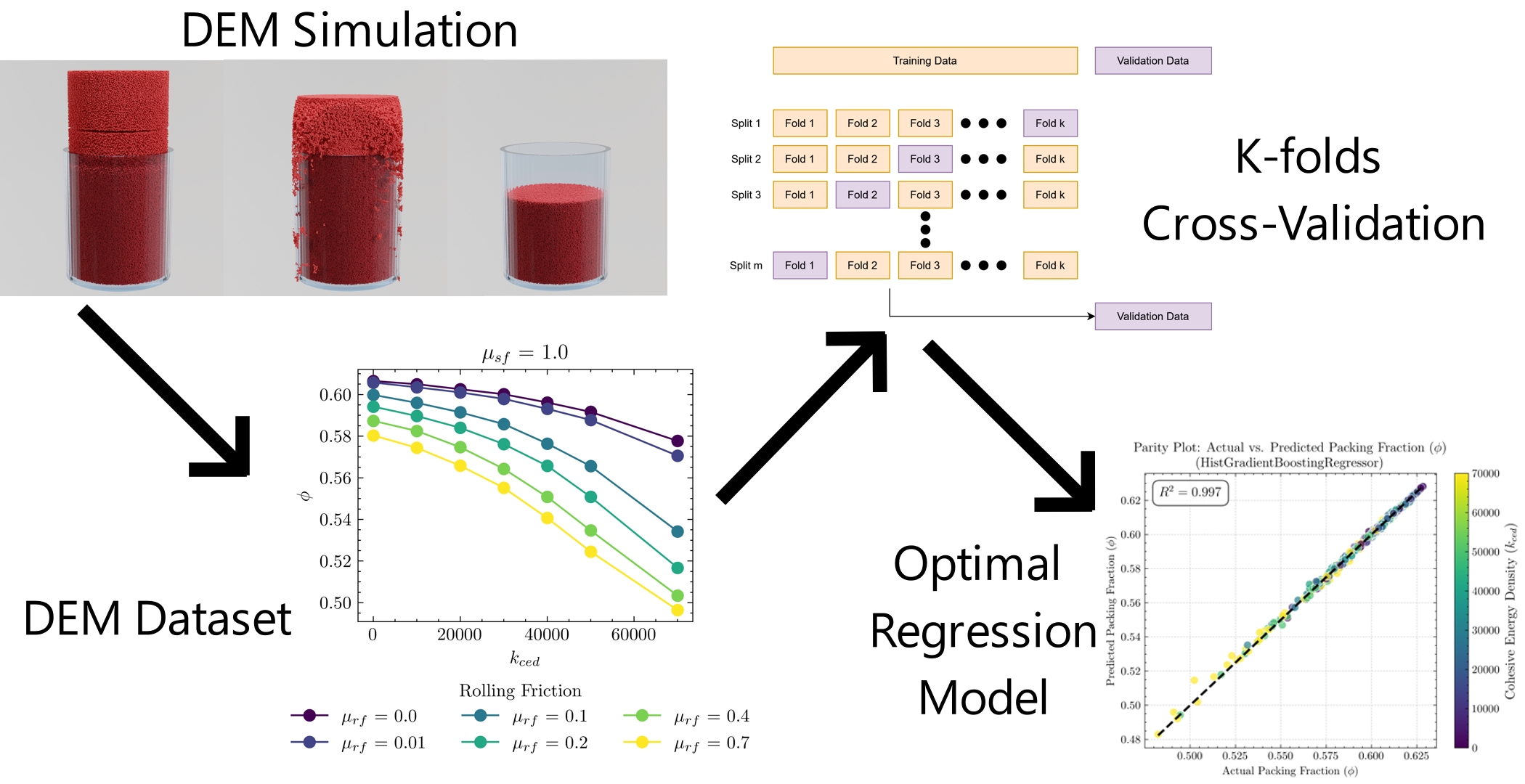}
\end{figure}
\end{graphicalabstract}

\begin{highlights}
\item A framework for benchmarking regression models for tabular DEM data is proposed.

\item Enables robust selection of fast and accurate surrogate models for DEM data.

\item Systematic model selection is crucial for effective DEM surrogate modelling.

\end{highlights}

\begin{keyword}
DEM Simulation \sep Machine Learning \sep Surrogate Modelling \sep Meta-modelling \sep K-fold Cross-validation

\end{keyword}

\end{frontmatter}



\begin{table}[h!]
\centering
\caption{Table of Symbols}
\begin{tabular}{@{}lll@{}}
\toprule
\textbf{Symbol} & \textbf{Description} & \textbf{Dimensions} \\
\midrule
$d_{50}$ & Median particle diameter & $L$ \\
$\epsilon$ & Coefficient of restitution & Dimensionless \\
$k_{ced}$ & Cohesive energy density& $M L^{-1} T^{-2}$ \\
$\mu_{rf}$ & Rolling friction coefficient & Dimensionless \\
$\mu_{sf}$ & Sliding friction coefficient & Dimensionless \\
$n$ & Number of samples & Dimensionless \\
$n_i$ & Number of particles of type $i$ & Dimensionless \\
$r_i$ & Radius of particle of type $i$ & $L$ \\
$R^2$ & Coefficient of determination & Dimensionless \\
$RMSE$ & Root Mean Square Error & $[y]$ \\
$MAE$ & Mean Absolute Error & $[y]$ \\
$V_p$ & Total particle volume & $L^3$ \\
$V_T$ & Total volume of the system & $L^3$ \\
$\phi$ & Packing fraction & Dimensionless \\
$y_i$ & Actual value of sample $i$ & $[y]$ \\
$\hat{y}_i$ & Predicted value of sample $i$ & $[y]$ \\
$\bar{y}$ & Mean of actual values & $[y]$ \\
\bottomrule
\end{tabular}
\end{table}

\section{Introduction}
\label{sec:intro}


As the Discrete Element Method (DEM) gains popularity \cite{Windows-Yule2016NumericalCheck}, it is increasingly important to develop regression models from DEM data to build metamodels for rapid evaluation in industrial applications. While DEM simulations can provide useful results, they are computationally expensive to run. This high cost drives the need for metamodels that have a significantly lower inference time, making them practical for real-world use.

Metamodels, also known as surrogate models, are simple models that describe another more complex model. In the context of DEM, a metamodel is often a regression model that has been trained on more computationally expensive DEM data and can be evaluated much faster. An example would be fitting a polynomial equation to the DEM data relating the geometry parameters of a hopper to the mass flow rate at the exit of that hopper where the polynomial is a metamodel obtained by fitting to a set of detailed DEM simulations \cite{Fransen2021ApplicationStudy}. Many applications of metamodels with DEM simulation data have been used extensively in previous literature; including areas such as material calibration and the design of bulk handling equipment \cite{Kalay2022MassMethod, Jadidi2024AnalysisApplications, Liao2021Image-basedMethods, Cui2024SimulationProcessing, Kumar2018PredictionNetwork, Richter2021IntroducingMethod, Irazabal2023ASurrogates, Rackl2017AModels,Hadi2025SystematicModels, Richter2020DevelopmentCalibration, BenTurkia2019BenefitsExperiments}, highlighting the importance of developing models with reduced evaluation time.

Regression models are also finding use in the calibration of DEM simulations, modelling the relationships between bulk measurements of powders and the microscopic particle interaction parameters needed for a material to be calibrated in a DEM simulation \cite{Boikov2018DEMForest, Rackl2017AModels, Benvenuti2016IdentificationExperiments, Fransen2022IncludingCalibration}.

Regression modelling is already widely used with DEM simulation data but little research has been done into which model is best and how to determine which model to use. In this paper, we set out a standard methodology for benchmarking a wide variety of models on DEM data from a data science perspective. An example use case of this methodology is conducted on a dataset of packing fraction data generated from DEM simulations. 

While the main aim of this paper is to provide a methodology for comparing the performance of regression models, the example use case of the packing fraction model presented here is a useful tool in its own right. It can be used to predict the packing fraction for a given set of DEM parameters, which is valuable when setting up simulations that require a set fill volume.

For example, achieving a specific fill level, such as 50\% in a rotating drum, is normally a trial and error process because the packing fraction of the simulated material is unknown beforehand. However, by using the regression model developed in this work, one can instantly predict the packing fraction of the material. This allows for the precise calculation of the number of particles needed to achieve the target fill volume on the first attempt, saving significant setup time.

\section{General Methodology}
\label{sec:methods}

\begin{figure}
    \centering
    \includegraphics[width=0.35\linewidth]{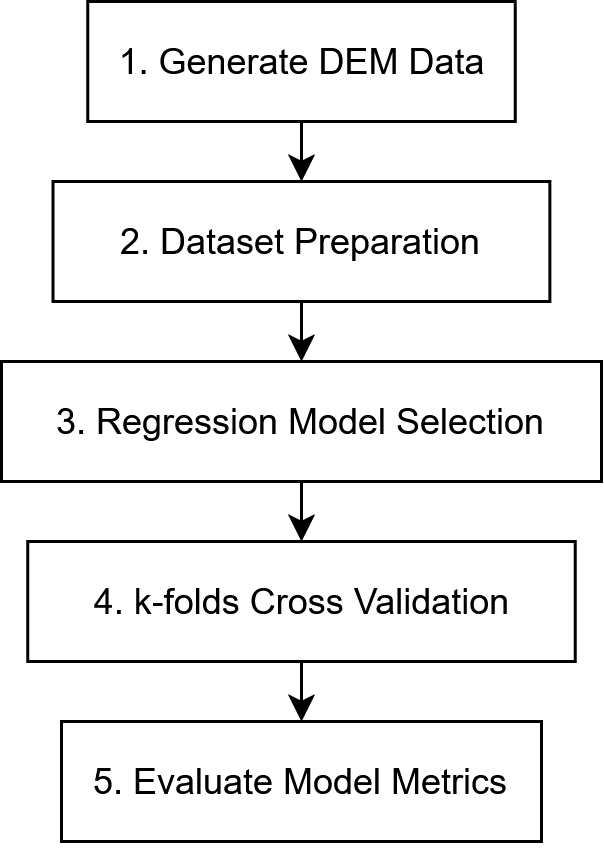}
    \caption{Diagram of the steps for benchmarking different regression models for DEM data.}
    \label{fig:benchmark_diagram}
\end{figure}

The general methodology for benchmarking regression models on DEM data in this study utilises k-fold cross-validation, a method of fairly comparing the performance of various models that will be explained in more detail in Section \ref{sec:methods}. This approach ensures accurate performance across the entire dataset and makes efficient use of data, which is especially useful for smaller datasets \cite{Abu-Mostafa2012LearningData}. Figure \ref{fig:benchmark_diagram} provides an overview of the five steps outlined in this framework for evaluating regression models on a DEM dataset.

\subsection{Generate DEM Dataset}


 Firstly, a comprehensive dataset is built that includes the relevant independent and dependent variables of interest for the target system's DEM simulations over the full range of parameters of interest. When doing so, it is important to consider the available computational resources and the time required to run all simulations, as well as the chosen design of experiments (e.g., full factorial or fractional factorial).

\subsection{Dataset Preparation}



One of the most important steps in regression model development is preprocessing the dataset. This ensures a high-quality input, which is essential for developing an accurate and reliable metamodel \cite{Zhang2003DataMining}. Proper preprocessing is not only critical for the final model's performance but also for the fair comparison of different models during the k-folds cross-validation step that will be conducted in Section \ref{sec:kfolds}. Data preprocessing typically improves data quality by removing erroneous points, reducing the dataset's size, and applying transformations. A brief introduction to this topic is presented below, along with some typical operations. More in-depth literature can be found in these references \cite{Zhang2003DataMining, Garcia2016BigProspects, Ridzuan2022DiagnosticAnalytics, Tukey1977ExploratoryAnalysis, Bolon-Canedo2015RecentData, Singh2020InvestigatingPerformance}.

\begin{enumerate}
    \item Removing Missing and Filtering Outlier Data
    \item Dimensionality Reduction
    \item Normalisation
\end{enumerate}

While using DEM simulations to generate a dataset greatly reduces the chance of missing values compared to experimental data collection, issues can still arise from simulations crashing or file corruption. In such cases, it is crucial that the post-processing step correctly identifies failed simulations and assigns them a null value. Subsequently, the regression model development must handle these missing values, typically by removing the corresponding data points before training. Missing data can substantially reduce training efficiency and introduce bias, thus impairing the model's accuracy \cite{Garcia2016BigProspects}.

Anomalous, outlier, and noisy data can also have a significant effect on model accuracy \cite{Garcia2016BigProspects}. Datasets from DEM simulations typically have little noise and few outliers because they provide direct access to exact particle conditions (e.g., positions, velocities). This eliminates potential measurement errors inherent in physical experiments. However, data points that are not useful to the final metamodel may still exist for various reasons (e.g., the granular material has entered a different flow regime that is not of interest). Outliers and noisy data can be identified for removal by applying noise filters or by using visualisation techniques, such as box plots and scatter plots, to visually inspect the data \cite{Garcia2016BigProspects, Tukey1977ExploratoryAnalysis}.

Dimensionality reduction is the process of reducing the number of dimensions—typically the input variables—of a dataset. This can be achieved through two main approaches: space transformation, which generates a smaller set of new features from a combination of the original ones, or feature selection, which removes irrelevant features from the dataset \cite{Bolon-Canedo2015RecentData}. A common space transformation method is principal component analysis (PCA), which creates new features as linear combinations of the original inputs. These new features are designed to explain the maximum possible variance in the output variable \cite{Dunteman1989PrincipalAnalysis}.

Feature selection improves data quality by removing irrelevant input features that do not contribute significant information, thereby simplifying the model without sacrificing accuracy \cite{Bolon-Canedo2015RecentData, Garcia2016BigProspects}. Numerous methods exist for feature selection. While they will not be covered here for brevity, Bolón-Canedo et al. \cite{Bolon-Canedo2015RecentData} provide a detailed exploration of both traditional and state-of-the-art techniques.

Normalisation  is the process of transforming the data values for each input variable (also known as a feature
) to a common, standardised scale. In the context of this study, features are the physical input parameters for the regression model, such as particle diameter, the coefficient of friction, or cohesion \cite{Singh2020InvestigatingPerformance, Garcia2016BigProspects}. For example, consider two input features with vastly different scales, such as $500-10,000$ and $10^{-8}-10^{-6}$. Without normalisation, many regression models would incorrectly assign greater importance to the first feature simply due to its larger magnitude. By scaling both features to a standard range, like 0 to 1, their initial magnitudes no longer disproportionately influence the model. Numerous normalisation techniques exist, and Singh and Singh \cite{Singh2020InvestigatingPerformance} provide a comprehensive overview.

\subsection{Regression Model Selection}

For benchmarking, a diverse suite of regression models with varying complexities is selected to identify the optimal model. A non-exhaustive list includes:
\begin{itemize}
    \item Linear models (e.g., Ordinary Least Squares, Ridge, Lasso).
    \item Non-linear models such as Polynomial Regression, Support Vector Machines (SVM), tree-based methods (e.g., Decision Trees, Random Forests, Gradient Boosting Machines), and Artificial Neural Networks (ANNs).
    \item Other relevant metamodelling techniques (e.g., Gaussian Process Regression, Symbolic Regression (for example; MED \cite{Nicusan2022PyMED:Discovery})).
\end{itemize}

\subsection{K-Fold Cross Validation}
\label{sec:kfolds}



\begin{figure}
    \centering
    \includegraphics[width=0.9\linewidth]{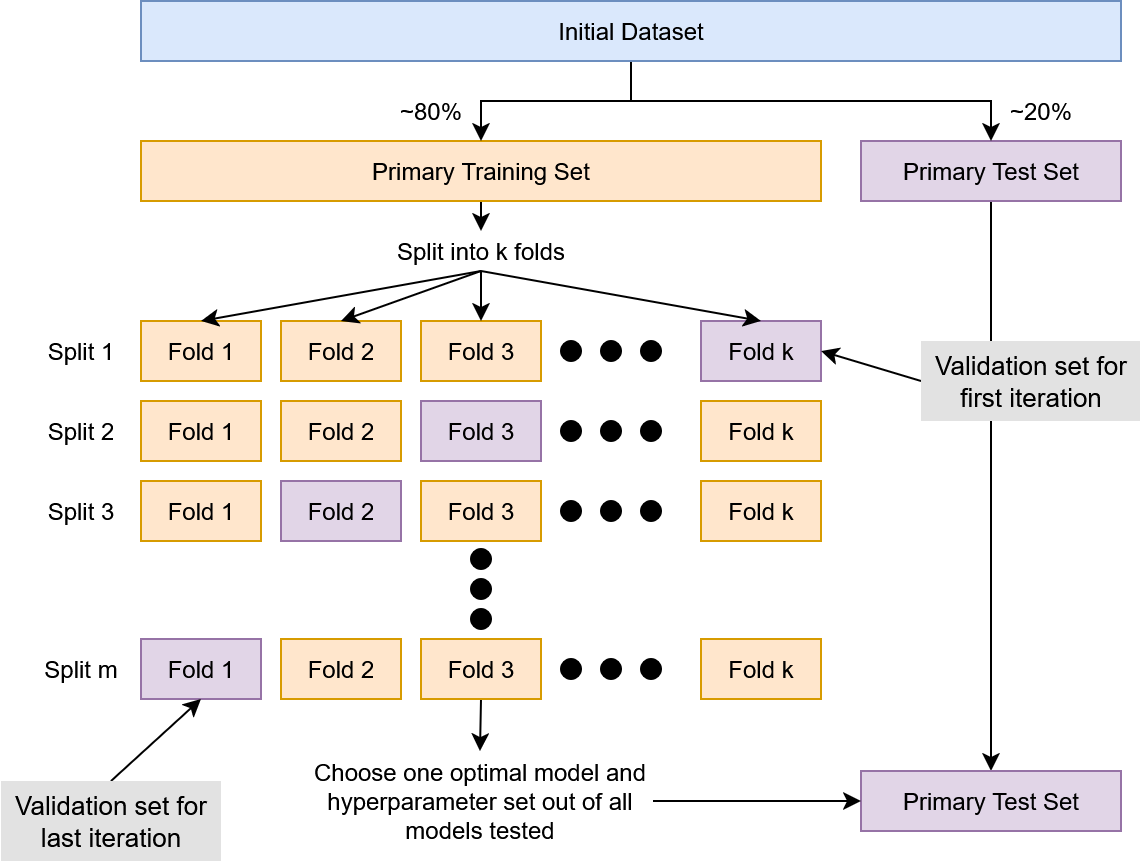}
    \caption{Diagram of k-fold cross validation process.}
    \label{fig:k_folds}
\end{figure}

K-fold cross-validation is implemented to provide reliable estimates of model generalisation performance \cite{Stone1974Cross-ValidatoryPredictions, Abu-Mostafa2012LearningData, James2013AnLearning}. Figure \ref{fig:k_folds} illustrates the steps involved in this process.

Initially, the entire dataset is split into an unseen primary training set and a primary test set. This test set is held in reserve and is only used at the very end to assess the final, selected model's performance. The primary training set is then divided into $k$ equally sized folds. For each of the m iterations (i.e., a single cycle of model training and validation), one fold serves as the validation set for that iteration, while the remaining $k-1$ folds are combined to form the training set. During each of these splits, model hyperparameters (the settings of the regression model, e.g., learning rate or number of hidden layers) are optimised using random search to ensure that each model performs optimally on its respective training folds. This process is repeated $m$ times, ensuring every fold has served as the validation set precisely once \cite{Stone1974Cross-ValidatoryPredictions, James2013AnLearning}.

After each split, the evaluation metrics of the trained model, as described in the next section, are recorded. The optimal number of folds to use, $k$, is debated but generally $k=5$ or $k=10$ has been found to provide good results \cite{Kohavi1995ASelection, James2013AnLearning}. An example Python script for k-folds cross validation can be found on GitHub: \href{https://github.com/BenDJenkins/K-Folds-Cross-Validation-Example}{https://github.com/BenDJenkins/K-Folds-Cross-Validation-Example}.


\subsection{Evaluate Model Metrics}

    It is important to consider a range of model metrics from performance metrics, that indicate how accurate a model is, to time metrics, that describe how long it takes to train and use a model. A few key metrics are discussed below.

    Coefficient of Determination ($R^2$): This metric represents the proportion of the variance in the dependent variable that is predictable from the independent variables. A value closer to 1 indicates a better fit, signifying that the model explains a larger portion of the variability in the data. Equation \ref{eqn:r_squared} shows the calculation of $R^2$.
    
    \begin{equation}
    \label{eqn:r_squared}
        R^2 = 1 - \frac{\sum_{i=1}^{n} (y_i - \hat{y}_i)^2}{\sum_{i=1}^{n} (y_i - \bar{y})^2}
    \end{equation}

    \noindent where $y_i$ is the actual observed value for each sample $i$, $\hat{y}_i$ is the corresponding value predicted by the model, $\bar{y}$ is the mean of all observed values, and $n$ is the total number of samples.

    Mean Absolute Error (MAE): MAE measures the average magnitude of the errors between the predicted and actual values, without considering their direction. It is given by Equation \ref{eqn:MAE}. A lower MAE indicates better performance, as it reflects smaller average prediction errors.

    \begin{equation}
    \label{eqn:MAE}
        MAE = \frac{1}{n} \sum_{i=1}^{n} |y_i - \hat{y}_i|
    \end{equation}

    \noindent where $n$ is the total number of samples, $y_i$ is the actual observed value, and $\hat{y}_i$ is the corresponding value predicted by the model.

    Root Mean Squared Error (RMSE): Similar to MAE, RMSE also quantifies the average magnitude of prediction errors. However, by squaring the errors before averaging, RMSE gives a higher weight to larger errors. It is calculated using Equation \ref{eqn:RMSE}. A lower RMSE is preferable.

    \begin{equation}
    \label{eqn:RMSE}
        RMSE = \sqrt{\frac{1}{n} \sum_{i=1}^{n} (y_i - \hat{y}_i)^2}
    \end{equation}

    \noindent where $n$ is the total number of samples, $y_i$ is the actual observed value, and $\hat{y}_i$ is the corresponding value predicted by the model.

   Many other possible metrics for evaluating model accuracy also exist such as Mean Bias and the Normalised Mean Error. Plevris et al. \cite{Plevris2022InvestigationModels} and Miller et al. \cite{Miller2024AGenomics} provide comprehensive overviews of other accuracy metrics. The coefficient of determination ($R^2$), mean average error (MAE) and root mean squared error (RMSE) were chosen to give a comprehensive overview of the model performance, exploring different aspects of the model performance (i.e. explained variance compared to average prediction error).


    Time metrics are also important to consider; two vital benchmarks are training time and inference time. Training time is the duration required for the model to learn from the training dataset. Inference time is the time taken by the trained model to make predictions based on some inputs. Both metrics can typically be measured using a given programming language’s built-in timing utilities or profiling tools.

    The evaluation metrics collected from each fold of the cross-validation process are aggregated to assess the performance of each model. Each model is compared against the others based on the averaged performance metrics and the standard deviation of the performance metrics. To compare where two models are statistically different, paired t-tests can be used. Additionally, computational aspects such as training time and inference time should be considered, especially since the goal is typically to develop models with reduced evaluation time compared to full DEM simulations. The final model selection should be based on the requirements of the problem.

\section{Example Methodology}
\label{sec:example}

    \subsection{Simulation Setup}
    \label{sec:sim_setup}

    The packing behaviour of granular materials is highly sensitive to inter-particle properties. This sensitivity is critical in applications where achieving a specific fill level, rather than just a total mass, is paramount. For instance, in mixing processes like Resonant Acoustic Mixing (RAM), the powder fill height can significantly influence system dynamics \cite{Sezer2025ExploringTracking}.

    To investigate these packing phenomena in a controlled yet relevant manner, this study utilises a simple beaker simulation. This system was chosen for its simplicity, which allows for a clear explanation of the example methodology, while still being an interesting real-world challenge of linking inter-particle properties to packing fraction. The beaker is a cylinder with the top open and the bottom capped off. The cylinder has an internal radius of 2~cm and a height of 5.5~cm, resulting in a total inner volume of 69.11~cm$^3$.

    The software used for these simulations is PICI-LIGGGHTS \cite{UniversityofBirminghamPositronImagingCentre2022UoBPICI-LIGGGHTS-3.8.1}, a modified version of the LIGGGHTS DEM engine \cite{Kloss2012ModelsCFD-DEM}. For computational efficiency, spheres are used in this study. Particle normal and tangential forces are calculated using the Hertz-Mindlin contact model \cite{Coetzee2017Review:Method, Hertz1882UeberKorper., Mindlin1953ElasticForces} in conjunction with Coulomb's law of friction to model the maximum tangential stress at which gross sliding occurs (sliding friction). To investigate how reduced particle rotation (due to rough or non-spherical particles) affects packing fraction, the constant directional torque (CDT) rolling resistance model \cite{Ai2011AssessmentSimulations} is employed to add an opposing rotational torque to the particles. Additionally, the simplified Johnson-Kendall-Roberts (SJKR) model \cite{Johnson1971SurfaceSolids, Hrvig2017OnSimulations} is used as a simple and computationally efficient contact model for particle cohesion. These contact models were chosen due to their wide-spread use and relatively low computational cost, striking a balance of accuracy and speed.
    
\begin{figure}
        \centering
        \includegraphics[width=1\linewidth]{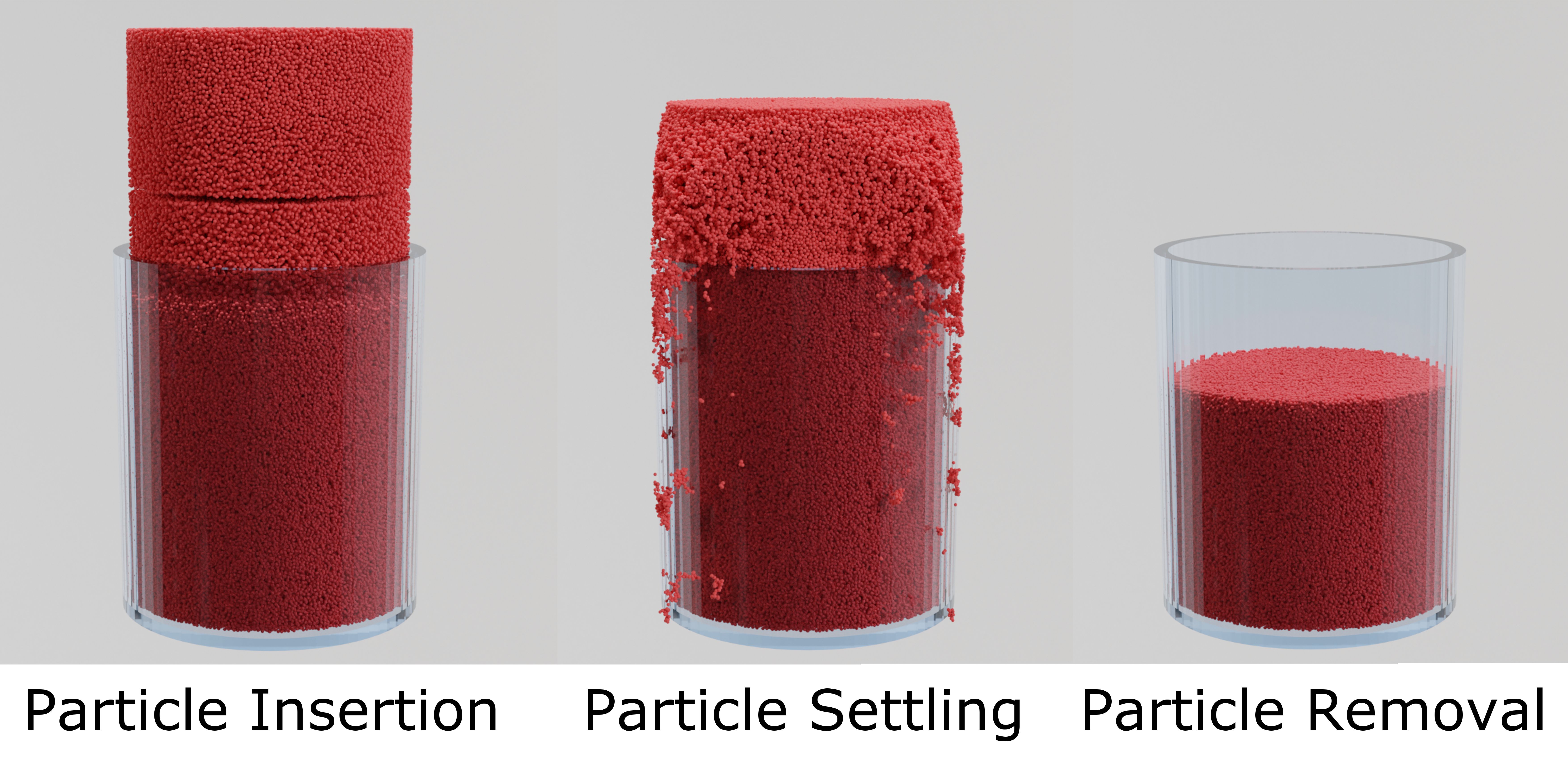}
        \caption{Steps of the LIGGGHTS beaker simulation.}
        \label{fig:beaker_steps}
    \end{figure}

    Particles are inserted into the simulation several centimetres above the open top of the cylinder, then allowed to fall into it, as depicted on the left of Figure \ref{fig:beaker_steps}. A gap is maintained between the top of the cylinder and the insertion region, allowing any overflowing particles to escape and avoid unwanted compaction. The simulation's objective is to measure how many particles fit into a 50 $cm^3$ volume within the cylinder (and thus the packing fraction $\phi$). To achieve this, an initial quantity of particles is inserted such that the total bulk volume of the material inside the cylinder will exceed 50 $cm^3$, as shown in the middle of Figure \ref{fig:beaker_steps}.

    First, particles are inserted into the beaker and allowed to settle (left and middle of Figure \ref{fig:beaker_steps}), the end point of which is marked by the system's total kinetic energy reaching a minimum. Next, to achieve a precise initial volume of 50~cm$^3$, any particles with a centre point above the corresponding fill height of 3.98~cm are removed from the simulation. This step is illustrated on the right in Figure \ref{fig:beaker_steps}. Finally, the remaining particles are then allowed to settle again to a minimal kinetic energy, and the deletion process is repeated to account for any bed expansion resulting from the removal of the overlying particle load.
    
    This cycle of deleting particles and allowing them to settle continues until fewer than 1\% of the particles are removed during a deletion step. For example, if there are 40,000 particles in the system, the simulation ends if fewer than 400 particles are removed in a single deletion step. The 1\% threshold was selected after testing over 100 simulations, ensuring accurate results without allowing the simulation to run indefinitely. Notably, for smaller particle diameters, the number of deleted particles often remained greater than zero, even after 10 deletion cycles.

    \subsection{Benchmarking Dataset}
    \label{sec:dataset}
    To create the benchmarking dataset, a full factorial design of experiments was used, leading to a total of 3024 beaker simulations that explored a wide range of particle properties. DEM simulations have a significant upfront cost to generate data for surrogate modelling due to their computational cost. For this example, the dataset size of 3024 simulations was chosen as a realistic target, representing approximately one month of data generation on a high performance computing (HPC) facility. Table \ref{tab:dem_sim_params} summarises both the constant DEM paramters such as the Young's modulus and material density across all the simulations as well as the range of $\epsilon$, $\mu_s$, $\mu_r$ and $k_{ced}$ values considered.

    Simulations were run on the University of Birmingham BlueBear HPC facility, using one core from an Intel Xeon Platinum 8360Y CPU per simulation. The runtime for each simulation ranged from three hours to nine days, primarily dependent on the particle $d_{50}$ due to the increased number of particles required at smaller diameters \cite{Windows-Yule2025NumericalSolutions}.


    \begin{table}[h!]
    \def\arraystretch{1.5}
    \centering
    \caption{Investigated and constant parameters for the beaker DEM simulations.}
    \label{tab:dem_sim_params}
        \begin{tabularx}{\linewidth}{ | >{\raggedright\arraybackslash}X || *{3}{>{\centering\arraybackslash}X|} }
        \hline
         \textbf{Simulation Parameter} & \textbf{$d_{50}=0.7\,\text{mm}$} & \textbf{$d_{50}=1.1\,\text{mm}$} & \textbf{$d_{50}=1.5\,\text{mm}$} \\
         \midrule\midrule
         Timestep [$seconds$] & $1.7e^{-6}$ & $6.3e^{-6}$ & $8.7e^{-6}$ \\
         \hline
         Young's Modulus ($E$) & \multicolumn{3}{c|}{$5e^6$} \\
         \hline
         Material Density [$kgm^{-3}$] & \multicolumn{3}{c|}{1000} \\
         \hline
         Coefficient of Restitution ($\epsilon$) & \multicolumn{3}{c|}{0.1, 0.5, 0.9, 0.99} \\
         \hline
         Sliding Friction ($\mu_s$) & \multicolumn{3}{c|}{0.15, 0.2, 0.3, 0.5, 0.8, 1.0} \\
         \hline
         Rolling Friction ($\mu_r$) & \multicolumn{3}{c|}{0.0, 0.01, 0.1, 0.2, 0.4, 0.7} \\
         \hline
         Cohesive Energy Density ($k$) [kJ/m$^3$] & \multicolumn{3}{c|}{0, 10000, 20000, 30000, 40000, 50000, 70000} \\
         \hline
        \end{tabularx}
    \end{table}
    
    After each simulation the volume of particles is calculated using Equation \ref{eqn:par_vol_total} and the packing fraction calculated using Equation \ref{eqn:packing_fraction}.

    \begin{equation}
    \label{eqn:par_vol_total}
        V_{\text{p}} = \sum_{i=1}^{N} n_i \cdot \left( \frac{4}{3} \pi r_i^3 \right)
    \end{equation}

    \noindent where $V_{\text{p}}$ is the total volume of all particles, $N$ is the number of distinct particle size class, $n_i$ is the number of particles in the $i$-th class, and $r_i$ is the radius of the particles in the $i$-th class.

    \begin{equation}
    \label{eqn:packing_fraction}
    \phi = \frac{V_p}{V_T}
    \end{equation}

    \noindent where $\phi$ is the packing fraction, $V_p$ is the total volume of the particles, and $V_T$ is the total volume of the system containing the particles.

\subsection{Overview of Regression Models}
\label{sec:regression_models}
    
    In this example study, we considered 16 different regression models in 6 distinct categories. These categories encompass a wide spectrum of algorithms, from simple linear baselines to complex, non-linear ensembles.

    The first category comprises linear models, which included standard Linear Regression (fit via ordinary least squares and Stochastic Gradient Descent), regularised variants such as Ridge, Lasso, and ElasticNet, and Partial Least Squares (PLS) Regression for handling collinearity. Filzmoser and Nordhausen \cite{Filzmoser2021RobustOverview} cover these and more linear regression models in detail.
    
    The foundational tree-based models formed the second category. This included decisions trees and random forests. A decision tree operates by recursively splitting data based on feature values to arrive at a prediction. While highly interpretable, a single tree can easily over-fit the training data. The random forest model addresses this limitation. It constructs a multitude of decision trees on various random subsets of the data and then aggregates their individual predictions (typically by averaging) to produce a single, more accurate value \cite{Grinsztajn2022WhyData}.
    
    Boosting ensembles are based on building models sequentially, where each new model corrects the errors of its predecessor \cite{Natekin2013GradientTutorial}. This category included the classic AdaBoost \cite{Schapire2013ExplainingAdaBoost} algorithm and several state-of-the-art gradient boosting implementations: Gradient Boosting \cite{Natekin2013GradientTutorial}, XGBoost \cite{Chen2016XGBoost}, LightGBM \cite{Ke2017LightGBM:Tree}, and HistGradientBoosting \cite{Natekin2013GradientTutorial}.
    
    Three more non-linear models were tested, each in a unique category. The K-Nearest Neighbors (KNN) Regressor, an instance-based method, works by predicting a value for a new data point based on the average of its k closest neighbors in the training set \cite{Song2017AnRegression}. The Support Vector Machine (SVM), a kernel-based method, works by finding an optimal hyperplane that fits the data while tolerating errors within a specified margin \cite{Cortes1995Support-vectorNetworks, Brereton2010SupportRegression}. Lastly, a Multi-layer Perceptron (MLP), a type of artificial neural network, uses interconnected layers of nodes to learn complex non-linear patterns by adjusting the weights between them during training \cite{Gardner1998ArtificialSciences}.

\section{Results and Discussion}
    \subsection{Dataset Generation}
    \label{sec:dataset_prep_example}

\begin{figure}
        \centering
        \includegraphics[width=0.95\linewidth]{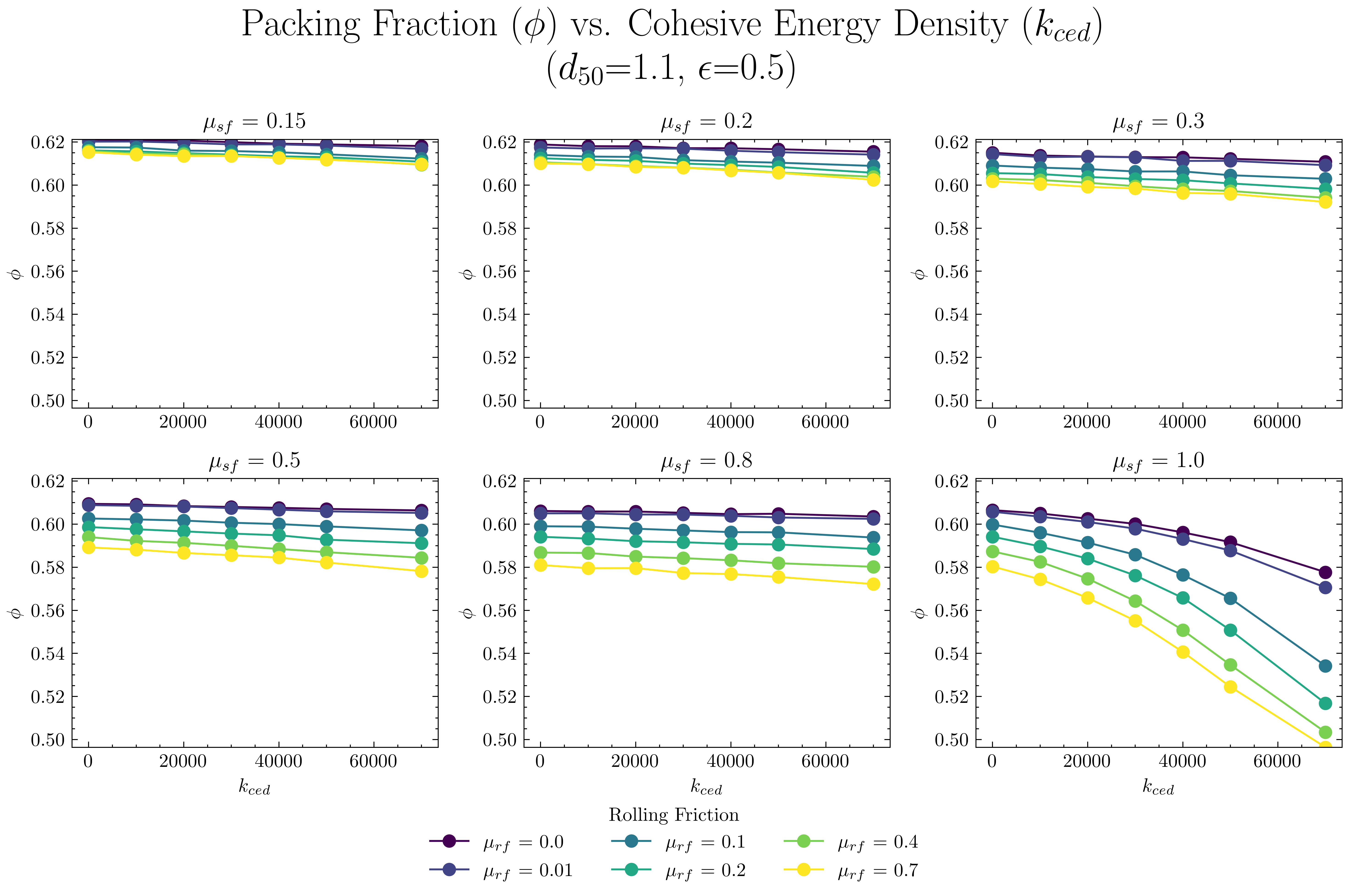}
        \caption{Packing fraction measured in the beaker simulation at different particle properties.}
        \label{fig:packing_fraction_scatter_plot}
    \end{figure}

    Once all simulations of the beaker across the parameter range set out in Section \ref{sec:dataset} were completed, the measured packing fraction results were aggregated and plotted on scatter graphs. These plots were then visually inspected to identify any anomalous data. An example scatter plot of the packing fraction data generated from the simulations is shown in Figure \ref{fig:packing_fraction_scatter_plot}. Due to the care taken in setting up the simulations and the extensive testing conducted before running the full study, no data points needed to be filtered out.

    As the dataset only contained four input features ($\epsilon, \, \mu_s, \, \mu_r, \, $and$ \, k_{ced}$) and all of them are important for the final model, no dimensionality reduction was conducted for this example dataset.

    The results were normalised using a Min-Max scaler, which rescales each feature to a specific range, in this case between 0 and 1. This provides a fair input into each of the regression models as some perform better with normalised data while others are not affected by it. Equation \ref{eqn:minmax} gives the equation for Min-Max normalisation.

    \begin{equation}
    	\label{eqn:minmax}
    	x_{\text{scaled}} = \frac{x - \min(X)}{\max(X) - \min(X)}
    \end{equation}
    
    \noindent where $x_{\text{scaled}}$ is the normalised value, $x$ is the original value, and $\min(X)$ and $\max(X)$ are the minimum and maximum values of the feature $X$, respectively.


    \subsection{Machine Learning Review}
    \label{sec:ml_review}

\begin{figure}
        \centering
        \includegraphics[width=0.98\linewidth]{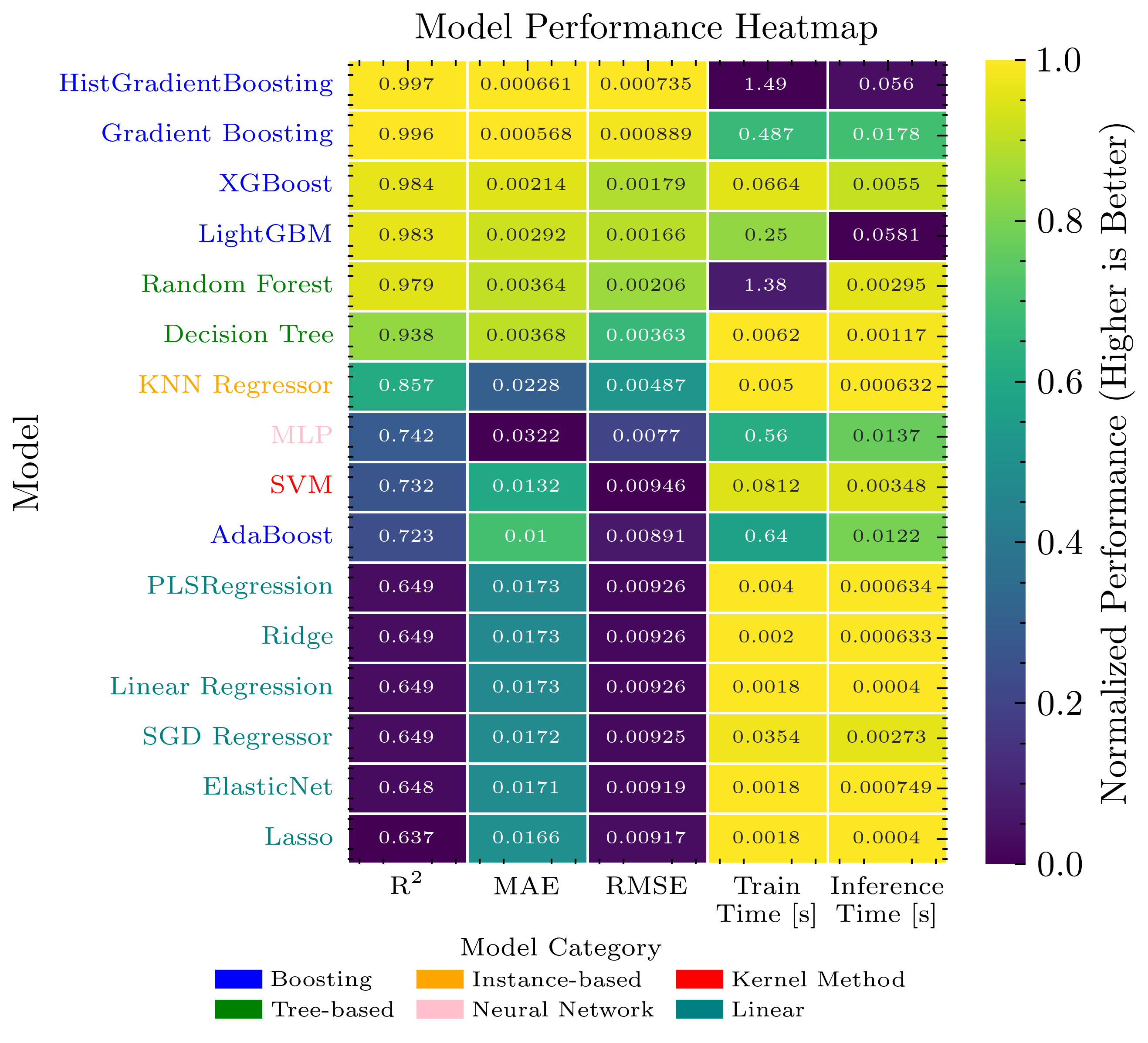}
        \caption{Heat-map results of k-folds cross validation on the 16 models tested.}
        \label{fig:regression_metrics}
    \end{figure}

    Figure \ref{fig:regression_metrics} summarises the 16 models discussed in Section \ref{sec:regression_models} in this study and their averaged performance metrics from the k-fold benchmark conducted on the packing fraction data, ordered by decreasing $R^2$ value. (A tabular version of the results can be found in \ref{sec:appendix_kfolds}). All models were implemented from and using the Python Scikit-learn library \cite{Pedregosa2011Scikit-learn:Python}. All training and testing were conducted using an Intel Core i5-2400 CPU.

    Overall, ensemble boosting methods performed the best. Notably, AdaBoost was the only boosting method to achieve a lower $R^2$ value than the non-boosting models. This is likely because, as one of the earliest practical boosting algorithms, its performance has since been surpassed by more modern methods like Gradient Boosting, which often achieve higher accuracy \cite{Schapire2013ExplainingAdaBoost}. The general success of boosting models here is expected, as they are known to perform well on tabular data for regression tasks \cite{Shwartz-Ziv2022TabularNeed, Rizkallah2025EnhancingProblems, Boldini2023PracticalPrediction, Grinsztajn2022WhyData}.

    Tree-based methods performed similarly to the top boosting models but were slightly less accurate across all accuracy metrics. This performance difference is expected, as gradient boosting models are ensembles of decision trees. This structure allows them to build upon the models in the sequence, progressively correcting the errors of the previous ones.
    
    Interestingly, the MLP (Multi-Layer Perceptron), an artificial neural network-based model, performed worse than the boosting and tree-based models. While its $R^2$ value was slightly higher than that of the linear models, its MAE was notably poorer highlighting the need to use various different metrics. This underperformance is likely due to the limited size of the dataset, as artificial neural networks typically require large amounts of data, typical 50 times more than the number of adjustable parameters (i.e. hyperparameters) \cite{Alwosheel2018IsAnalysis}. Furthermore, MLPs often have a larger number of hyperparameters to tune and can be more sensitive to their configuration compared to the other models tested.


    The HistGradientBoosting model, a histogram-based gradient boosting model, provided the best $R^2$, MAE, and RMSE values. However, it also took the longest time to train and had one of the longest inference times. Despite its longer training time, this was still on the order of seconds, which is acceptable for this model's use case. The HistGradientBoosting model was thus chosen due to its superior performance and acceptable training and inference times.



    \subsection{Final Model Training}
    \label{sec:final_model}

    Now the HistGradientBoosting model has been shown to be optimal, it will be retrained on the full training dataset (as depicted in orange at the top of the diagram in Figure \ref{fig:benchmark_diagram}). A final hyperparameter optimisation is then done using Optuna \cite{Akiba2019Optuna}. Finally, the retrained model will be evaluated on the initially held-out primary test set, which the model has not previously encountered. This final evaluation is a good check of the model's generalisability to unseen data.

\subsection{Final Model Evaluation}
\label{sec:results}

\begin{figure}
        \centering
        \includegraphics[width=0.7\linewidth]{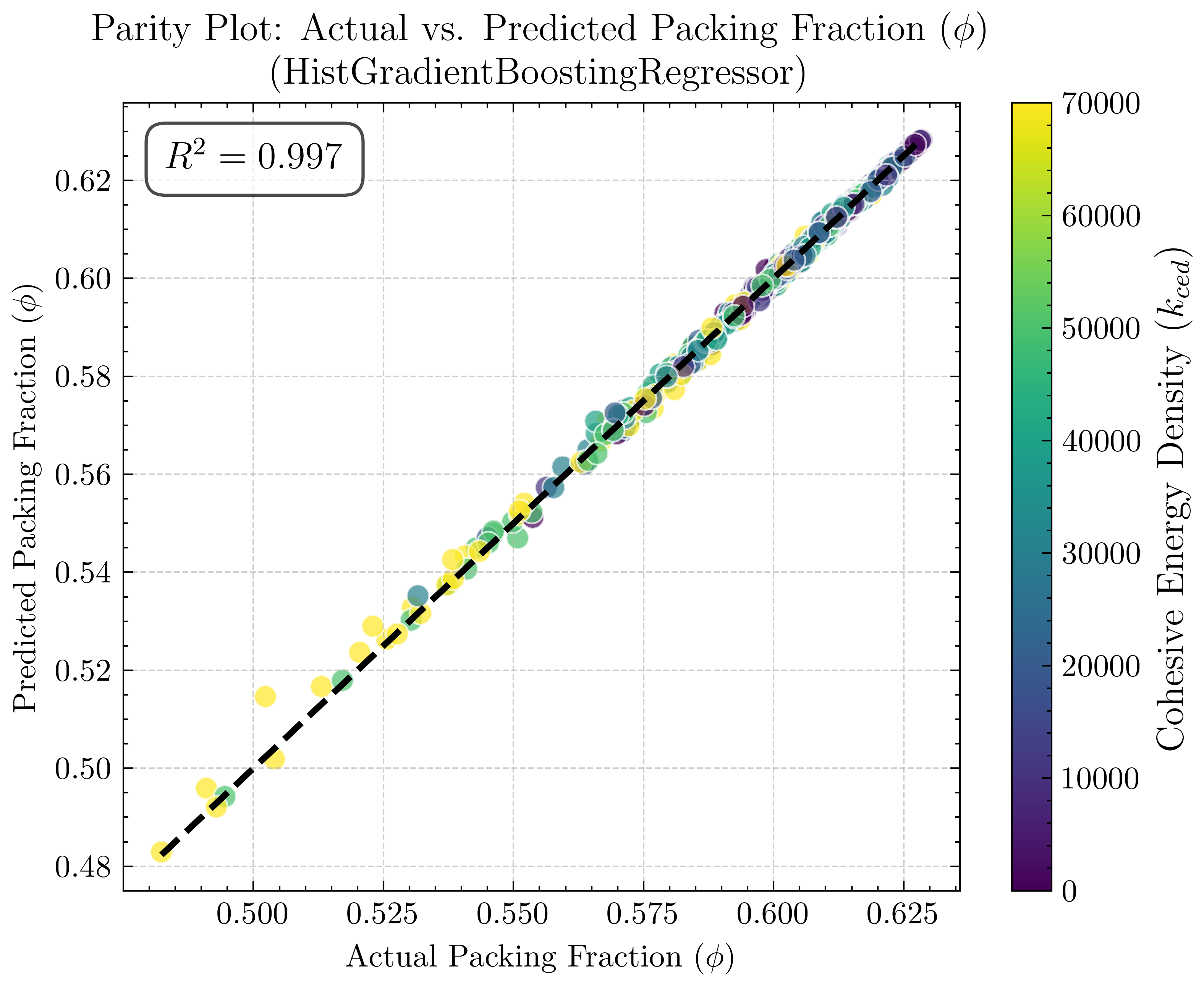}
        \caption{Parity plot of the unseen validation dataset for the HistGradientBoosting model.}
        \label{fig:parity_plot}
    \end{figure}

    Figure \ref{fig:parity_plot} shows a parity plot of the HistGradientBoosting model predicting the completely unseen validation dataset. Parity plots compare the predicted model output to the actual value measured for the same input values. The more closely the points lie to the diagonal, the more accurately the model predicts the data. Figure \ref{fig:parity_plot} shows that the trained HistGradianetBoosting model is able to accurately predict the packing fraction of the materials in the validation dataset. There are a few values at lower packing fractions where the model over-predicts the packing fraction as can be seen by the points that lie above the diagonal but overall the model is good from the results of the parity plot.

    SHAP (SHapley Additive exPlanations) plots are used to explain the impact of input variables on model predictions \cite{Lundberg2017APredictions}. This method is based on Shapley values, a concept from cooperative game theory. It calculates the marginal contribution of each feature toward an individual prediction, essentially assigning credit fairly among all features.

    Specifically, a feature's SHAP value quantifies exactly how much that feature's value shifted a single prediction away from the average prediction over the entire dataset. On a SHAP summary plot, features are ranked by their overall importance (from top to bottom). The plot then shows whether a particular input value increased (a positive SHAP value) or decreased (a negative SHAP value) the model's final output relative to this average \cite{Lundberg2017APredictions}.

\begin{figure}
        \centering
        \includegraphics[width=0.7\linewidth]{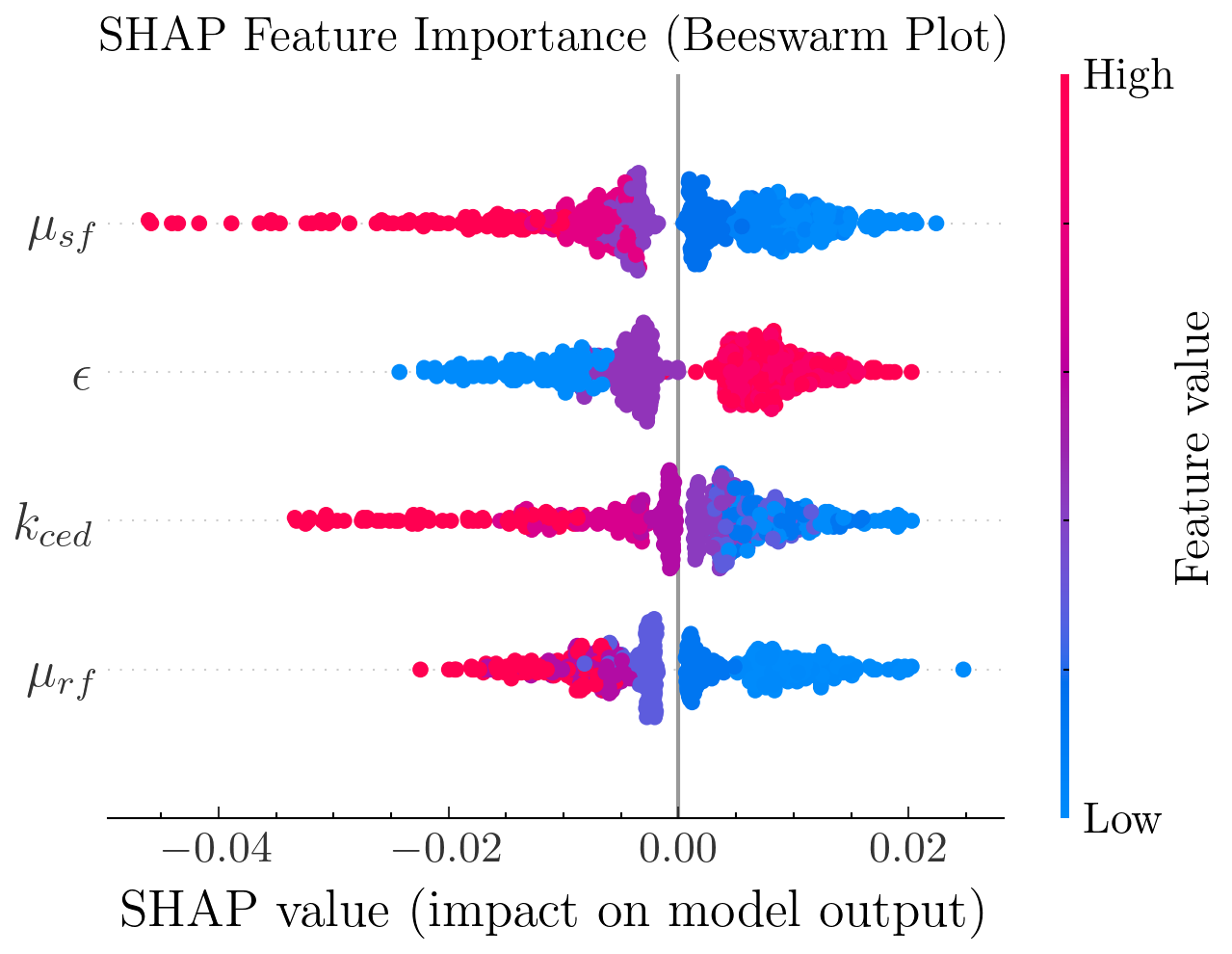}
        \caption{SHAP plot of the HistGradientBoosting model.}
        \label{fig:shap}
    \end{figure}

    For example, in Figure \ref{fig:shap}, the SHAP plot of the HistGradientBoosting model is presented. The most important input feature was the sliding friction ($\mu_{sf}$) at the top, and the least important was the rolling friction ($\mu_{rf}$) at the bottom. The blue points (lower values) for $\mu_{sf}$ have positive SHAP values, indicating that in the model, lower values of $\mu_{sf}$ increased the predicted packing fraction. Lower friction leading to a higher packing fraction makes intuitive sense: if particles can slide over each other more easily, they rearrange more easily and thus pack more efficiently.  This rearrangement is highly related to well established packing metrics such as the Hausner Ratio which quantifies the ability of a powder to rearrange itself based on a specific energy input (i.e., the difference between its bulk and tapped densities) \cite{Saker2019PowdersRatio}.

    The next most important feature in the model is the coefficient of restitution ($\epsilon$). This may appear surprising, but it can be understood by considering the particle settling process.
    

    The effect of the coefficient of restitution can be thought of like particle deposition in filtration. Systems with low kinetic energy and high inter-particle cohesion tend to form open, porous structures, as particles stick upon first contact. Conversely, systems with high kinetic energy and low attraction behave like a ``pin-ball machine'' where particles can rattle around and settle into more efficient, densely packed arrangements \cite{Rhodes2024IntroductionTechnology, Seville2016MechanicsSolids}.

    In the simulations, particles are introduced by dropping them into the container. A higher coefficient of restitution better preserves the particles' kinetic energy after impacts with the container and other particles. This sustained energy allows for more extensive rearrangement and exploration of void spaces before the system comes to rest, ultimately leading to a higher final packing fraction.
    
    
    After $\epsilon$, $k_{ced}$ and $\mu_{rf}$ had similar effects on the packing fraction in the model as $\mu_{sf}$, with smaller values having positive SHAP values, indicating they contributed to predictions of higher packing fraction. This again intuitively makes sense, as both resistance to rotation and stronger cohesion hinder particles from moving past each other and rearranging into a more efficient packing.



\section{Conclusion and Recommendations}
    
    This work demonstrated a k-fold cross-validation framework for benchmarking regression models on DEM datasets, addressing a common gap where model selection is often not considered. Applying this framework to a packing fraction dataset, a histogram-based gradient boosting model was identified as optimal due to its high performance and acceptable speed. It is the authors hope that this methodology will encourage more robust and data-specific model selection in future DEM studies.

    For researchers developing regression models from similar tabular DEM datasets, we strongly recommend considering ensemble boosting methods. Models such as HistGradientBoosting, Gradient Boosting, and XGBoost, delivered the highest accuracy in the benchmark. XGBoost, in particular, offers an excellent balance of high performance and low computational cost, making it a powerful and practical first choice for large datasets.

    While boosting models were superior, tree-based models like Random Forest also performed well and can be considered a simpler alternative where model interpretability is a priority. In contrast, other models were less suitable for this dataset's non-linear nature. Linear regression performed poorly, and while the Multi-Layer Perceptron (MLP) showed middling accuracy ($R^2$), its high mean average error (MAE) suggests it is not a good fit without a significantly larger dataset or better hyperparameter tuning.

\section*{Acknowledgements}

Authors acknowledge financial support received from the Centre for Doctoral Training in Formulation Engineering (EPSRC grant number EP/S023070/1) and Granutools. Computational resources have been provided by the University of Birmingham BlueBear facility (see http://www.birmingham.ac.uk/bear for more details).

\bibliography{references, references-2}

\appendix
\section{K-Folds Cross Validation Full Results}
\label{sec:appendix_kfolds}

    \begin{table}[h!]
    \centering
    \caption{Performance metrics for evaluated regression models.}
    \label{tab:regression_metrics}
    \renewcommand{\arraystretch}{1.2}
    \begin{tabular}{@{} p{3.7cm} p{1.1cm} p{1.3cm} p{1.3cm} p{1.2cm} p{1.8cm} @{}}
        \toprule
        \textbf{Model} & \textbf{R$^2$} & \textbf{MAE} & \textbf{RMSE} & \textbf{Train Time [s]} & \textbf{Inference Time [s]} \\
        \midrule
        \textcolor{blue}{HistGradientBoosting}     & 0.997 & 0.000661 & 0.000735 & 1.49 & 0.056 \\
        \textcolor{blue}{Gradient Boosting}        & 0.996 & 0.000568 & 0.000889 & 0.487 & 0.0178 \\
        \textcolor{blue}{XGBoost}                  & 0.984 & 0.00214 & 0.00179 & 0.0664 & 0.0055 \\
        \textcolor{blue}{LightGBM}                 & 0.983 & 0.00292 & 0.00166 & 0.25 & 0.0581 \\
        \textcolor{green!50!black}{Random Forest}  & 0.979 & 0.00364 & 0.00206 & 1.38 & 0.00295 \\
        \textcolor{green!50!black}{Decision Tree}  & 0.938 & 0.00368 & 0.00363 & 0.0062 & 0.00117 \\
        \textcolor{orange!80!black}{KNN Regressor} & 0.857 & 0.0228 & 0.00487 & 0.005 & 0.000632 \\
        \textcolor{pink}{MLP}                    & 0.742 & 0.0322 & 0.0077 & 0.56 & 0.0137 \\
        \textcolor{red}{SVM}    & 0.732 & 0.0132 & 0.00946 & 0.0812 & 0.00348 \\
        \textcolor{blue}{AdaBoost}                 & 0.723 & 0.01 & 0.00891 & 0.64 & 0.0122 \\
        \textcolor{teal!60!black}{PLSRegression}   & 0.649 & 0.0173 & 0.00926 & 0.004 & 0.000634 \\
        \textcolor{teal!60!black}{Ridge}           & 0.649 & 0.0173 & 0.00926 & 0.002 & 0.000633 \\
        \textcolor{teal!60!black}{Linear Regression} & 0.649 & 0.0173 & 0.00926 & 0.0018 & 0.000400 \\
        \textcolor{teal!60!black}{SGD Regressor}   & 0.649 & 0.0172 & 0.00925 & 0.0354 & 0.00273 \\
        \textcolor{teal!60!black}{ElasticNet}      & 0.648 & 0.0171 & 0.00919 & 0.0018 & 0.000749 \\
        \textcolor{teal!60!black}{Lasso}           & 0.637 & 0.0166 & 0.00917 & 0.0018 & 0.0004 \\
        \bottomrule
    \end{tabular}
    
    \vspace{0.5em}
    \textbf{Legend:}
    
    \begin{tabular}{@{} l l l @{}}
    \textcolor{blue}{\textbf{Blue}}: Boosting models &
    \textcolor{green!50!black}{\textbf{Green}}: Tree-based models &
    \textcolor{orange!80!black}{\textbf{Orange}}: Instance-based \\
    \textcolor{pink}{\textbf{Pink}}: Neural networks &
    \textcolor{red}{\textbf{Red}}: Kernel methods &
    \textcolor{teal!60!black}{\textbf{Teal}}: Linear models \\
    \end{tabular}
    \end{table}



\end{document}